# Context-Aware Hybrid Routing in Bluetooth Mesh Networks Using Multi-Model Machine Learning and AODV Fallback

Md Sajid Islam [1] and Tanvir Hasan [2]

Department of Artificial Intelligence and Big Data, Woosong University, Daejeon 34606, South Korea.
[1] mister.sajidislam@gmail.com
[2] tanvirhasan2427@gmail.com

## ABSTRACT

Bluetooth-based mesh networks offer a promising infrastructure for offline communication in emergency and resource-constrained scenarios. However, traditional routing strategies such as Ad hoc On-Demand Distance Vector (AODV) often degrade under congestion and dynamic topological changes. This study proposes a hybrid intelligent routing framework that augments AODV with supervised machine learning to improve next-hop selection under varied network constraints. The framework integrates four predictive models—a delivery success classifier, a TTL regressor, a delay regressor, and a forwarder suitability classifier—into a unified scoring mechanism that dynamically ranks neighbors during multi-hop message transmission. A simulation environment with stationary node deployments was developed, incorporating buffer constraints and device heterogeneity to evaluate three strategies: baseline AODV, a partial hybrid ML model (ABC), and the full hybrid ML model (ABCD). Across ten scenarios, the Hybrid ABCD model achieves ~99.97% packet delivery under these controlled conditions, significantly outperforming both the baseline and intermediate approaches. The results demonstrate that lightweight, explainable machine learning models can enhance routing reliability and adaptability in Bluetooth mesh networks, particularly in infrastructure-less environments where delivery success is prioritized over latency constraints.

## INTRODUCTION

The proliferation of mobile and low-power devices has catalyzed the development of Bluetooth Low Energy (BLE) mesh networks, which offer a promising infrastructure for communication in environments lacking stable internet connectivity. These networks enable device-to-device message relaying in real time, forming ad hoc topologies suitable for scenarios such as disaster recovery, crisis communications, and smart infrastructure monitoring. BLE mesh networking extends the core BLE stack to support many-to-many communication, enabling the creation of large-scale networks with thousands of nodes operating collaboratively [1].

Despite these advantages, maintaining reliable multi-hop routing in BLE mesh deployments presents unique challenges due to unpredictable link variability, limited buffer capacity, constrained energy resources, and restricted processing power at devices [2][3]. Traditional routing protocols like Ad hoc On-Demand Distance Vector (AODV) are designed to operate reactively by initiating route discovery only when needed. While this helps conserve resources, it may also introduce inefficiencies when unstable connections or congested buffers are selected as intermediate hops, leading to premature TTL expiry, message drops, or excessive routing latency [3][4][5]. These limitations are further amplified in BLE mesh deployments where physical range, buffer capacity, and link quality vary significantly across devices [6][7][8].

To address these challenges, researchers have explored various enhancements to traditional routing protocols. For instance, the Machine Learning-Based Optimized Routing Algorithm (ML-ORA) integrates parameter settings, a Hybrid Particle Swarm Optimization algorithm for cluster head selection, and a k-Nearest Neighbors-based intrusion detection system to offer adaptable and intelligent routing solutions in MANETs [9]. Similarly, reinforcement learning approaches have been applied to improve routing decisions by enabling nodes to learn optimal paths based on network dynamics [10].

In the context of Delay-Tolerant Networks (DTNs), machine learning techniques have been employed to enhance routing decisions. For example, a Multi-Decision Dynamic Intelligent (MDDI) routing protocol based on double Q-learning considers node relationships and message attributes to achieve efficient message transmission in DTNs [11]. However, these approaches often require extensive training data and may lack the flexibility to adapt to diverse operating conditions.

While these solutions offer improvements, they often focus on specific aspects of routing and may not provide a lightweight approach that balances reliability, efficiency, and interpretability in BLE mesh networks.

This work presents a hybrid routing framework that augments AODV with a set of lightweight, interpretable supervised machine learning models trained to predict delivery success, TTL usage, delay, and forwarder suitability. Each model operates in a fully decentralized manner, using only locally observable metadata from nodes and messages. A composite scoring function combines the outputs of these models to guide next-hop selection, while fallback to classical AODV ensures robustness under uncertain conditions.

Extensive simulations across stationary scenarios demonstrate that the proposed approach improves delivery reliability and resource efficiency under constrained conditions. While related work has explored ML-based routing and reinforcement learning, our framework uniquely combines multiple supervised models with context-aware scoring and AODV fallback to support scalable, real-time decision-making. Although BLE mesh often supports mobile devices, this study evaluates stationary deployments only, to provide a controlled feasibility baseline.

The remainder of the paper is structured as follows. Section 2 reviews related works on ML-based routing and hybrid decision-making in ad hoc networks. Section 3 outlines the proposed methodology, including the simulation framework, feature engineering, machine learning models, and model fusion logic. Section 4 presents the experimental setup, results, and an in-depth discussion of performance and limitations. Finally, Section 5 concludes the paper.

1. **RELATED WORK**

The integration of machine learning (ML) and intelligent decision-making into ad hoc routing protocols has been extensively explored in recent literature, with particular focus on enhancing delivery reliability, minimizing latency, and securing mobile networks. However, most prior works either optimize a single objective (e.g., delivery success or intrusion detection), rely on resource-intensive learning techniques (e.g., deep reinforcement learning or federated learning), or omit fallback mechanisms essential for robustness in real-world deployments. This section reviews recent studies from 2023 to 2025, highlighting their methods, performance, and limitations, and positioning our proposed Hybrid ABCD ML-AODV model as a more comprehensive alternative.

Ali et al. (2023) proposed a secure routing enhancement to AODV using support vector machines (SVM) and artificial neural networks (ANN) to detect and isolate blackhole attackers in MANETs [12]. While their approach achieved a packet delivery ratio (PDR) of 97.96% and a low delay of 0.04s, it focused solely on malicious behavior detection. In contrast, our model emphasizes multi-hop delivery success, route efficiency, and TTL preservation across benign, resource-constrained networks, making it more broadly applicable.

Ibrahim and Ghanim (2024) reviewed a wide range of AI models used to secure AODV against wormhole and blackhole attacks [13]. Although they report high detection accuracy (up to 99.1%) and PDRs under certain attack models, the surveyed works primarily enhance security rather than improve real-time, decentralized routing performance. Our work, by comparison, delivers both high reliability (99.97% delivery success) and context-aware forwarding under controlled operating conditions.

Hanif et al. (2024) introduced a deep reinforcement learning (DRL) routing framework combined with blockchain for route authentication [14]. Their method achieved approximately 94% PDR and 72ms delay. However, DRL's convergence latency and computation cost pose challenges for real-time, low-power Bluetooth mesh environments. Our supervised ML framework avoids online training while achieving significantly higher PDR with acceptable delay, aided by a lightweight fallback mechanism.

Sharma et al. (2023) applied Q-learning to develop adaptive routing for VANETs, attaining around 95% PDR and low latency (~0.1s) [15]. While effective, their method is sensitive to topology fluctuations and does not account for buffer congestion or energy constraints. Our model uses engineered features including TTL, buffer load, uptime, and node role to make more informed decisions.

Fernández et al. (2024) implemented a federated learning-based approach to train context-aware routing models across decentralized nodes [16]. Though privacy-preserving and scalable, federated learning introduces periodic communication overhead and lacks explicit route failure recovery. Our approach maintains local-only inference and integrates a classical AODV fallback mechanism to preserve routing reliability.

Other approaches, such as energy-aware routing using fuzzy logic [17], demonstrated improvements in energy preservation but offered lower delivery rates (~90%) and lacked predictive adaptability. Unlike these heuristic models, our ABCD framework fuses four supervised models—each optimized for delivery likelihood, TTL usage, delay estimation, and neighbor ranking—into a unified scoring mechanism that responds intelligently to diverse contextual factors.

Table 1 Summary of Related Works

| Study | Methodology | Key Contributions | Results | Limitations |
| --- | --- | --- | --- | --- |
| Ali et al. (2023) | SVM + ANN (Supervised) | ML-based detection of blackhole attacks in MANETs | PDR: 97.96%, Delay: 0.04s | Security-focused only; no TTL/delay optimization |
| Ibrahim & Ghanim (2024) | AI Model Survey | Comprehensive survey on AODV attack detection methods | Detection Acc: ~99.1% | No routing logic or end-to-end performance analysis |
| Hanif et al. (2024) | DRL + Blockchain | Secure adaptive routing using deep Q-learning | PDR: ~94%, Delay: ~72ms | High cost; no TTL optimization; lacks fallback |
| Sharma et al. (2023) | Q-Learning (RL) | Adaptive VANET routing under mobility | PDR: ~95%, Delay: ~100ms | No TTL or buffer awareness; not BLE-oriented |
| Fernández et al. (2024) | Federated Learning (SVM) | Privacy-aware routing using edge training | PDR: ~96%, Delay: ~100ms | Model exchange overhead; no fallback logic |
| Faruqi et al. (2023) | Fuzzy Rule-Based | Energy-aware routing for mobile IoT | PDR: ~90% | No learning; lacks delay or delivery optimization |
| Proposed Framework (2025) | Multi-Model ML + AODV Fallback | 4-model ML fusion with delivery, TTL, delay prediction + fallback | PDR: 99.97%, Delay: 140ms | Slightly higher delay; needs real-world testing |

The proposed Hybrid ABCD model represents a novel integration of lightweight supervised learning models with a dynamic scoring function and AODV-based fallback. Unlike previous works that rely on a single learning model or focus exclusively on security, our approach offers real-time, context-aware routing with superior delivery success and TTL efficiency. By combining interpretable features with decentralized decision-making, it provides a robust and scalable solution for stationary Bluetooth mesh networks under constrained resources.

2. **PROPOSED METHODOLOGY**

The proposed methodology presents a hybrid intelligent routing framework designed for Bluetooth-based offline mesh networks in crisis scenarios. The primary objective is to achieve high message delivery success and low latency under conditions of device heterogeneity and communication constraints. The system integrates classical Ad-hoc On-Demand Distance Vector (AODV) routing with a multi-model machine learning (ML) strategy that intelligently selects next-hop neighbors during message delivery. Four predictive models—focused on delivery success probability, TTL cost, delay estimation, and forwarder ranking—are trained using synthetically generated datasets derived from AODV-based simulations. Contextual features such as buffer ratio, signal quality, uptime, priority tolerance, and distance to target are dynamically extracted in real time to support adaptive decision-making. The fusion of model outputs ensures robust and explainable routing behavior.

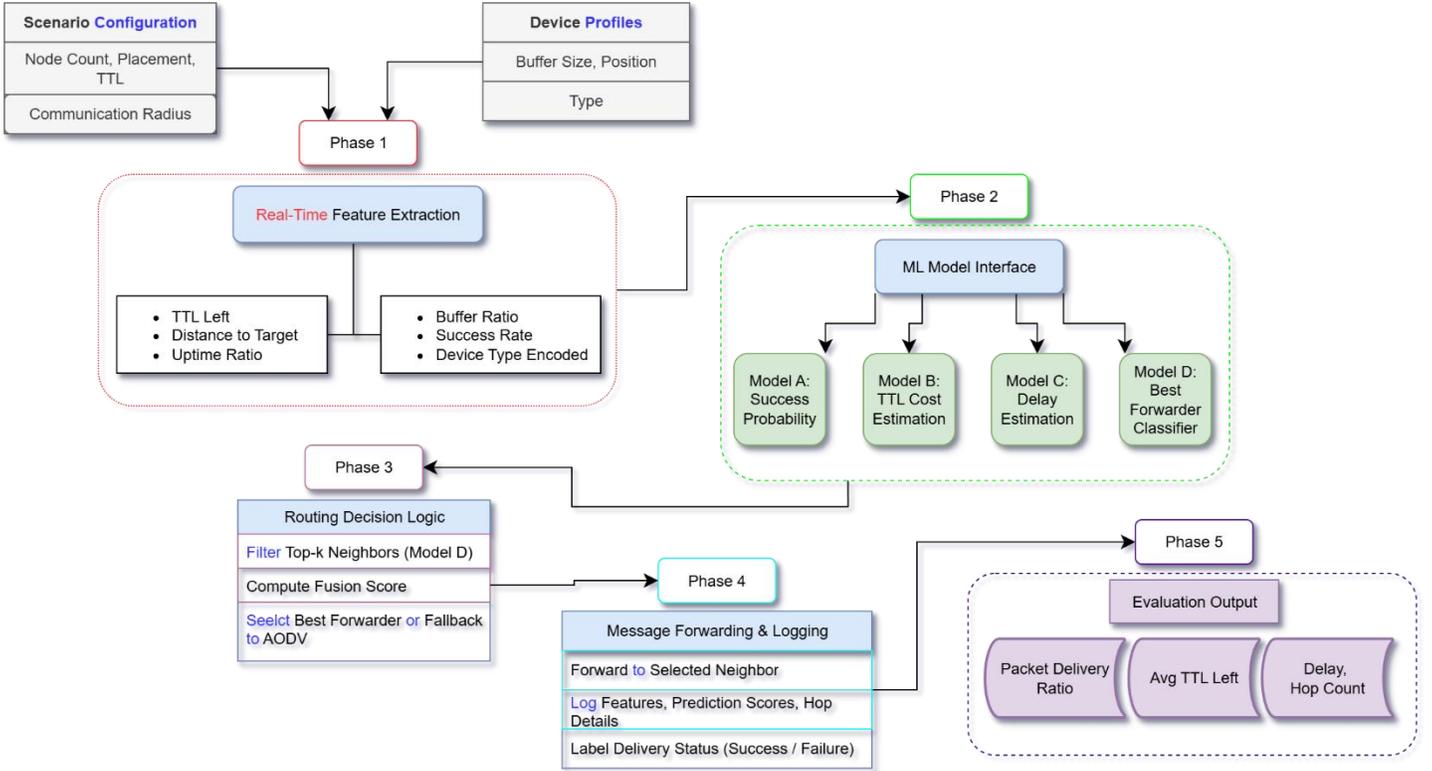

**Figure 1.** Proposed Framework of Context-Aware Hybrid Routing in Bluetooth Mesh Networks

Figure 1. Proposed Hybrid ABCD routing framework for Bluetooth mesh networks. The framework consists of five phases: (1) scenario and device initialization, (2) feature extraction (e.g., TTL left, buffer ratio, distance to target), (3) ML model predictions for delivery success, TTL cost, delay, and forwarder suitability, (4) routing decision using composite ABCD scoring with AODV fallback and message forwarding with hop-level logging, and (5) aggregation of performance metrics including PDR, TTL utilization, and delay. All evaluations in this study are conducted under stationary node deployments.

The following subsections describe each stage of the methodology, emphasizing the technical design and implementation of the system.

### 2.1. DATASETS

This study employs synthetically generated datasets that simulate Bluetooth mesh networks with stationary node deployments, incorporating device heterogeneity and network constraints. The datasets are designed to support both supervised machine learning training and robust performance evaluation in decentralized, ad hoc environments [18]. Each simulation begins with a scenario definition file (e.g., devices_scenario_1.csv) specifying node-level parameters including spatial position, buffer capacity, battery level, device type, and context-aware attributes such as uptime ratio and priority tolerance. These files model real-world deployment conditions such as heterogeneous IoT devices or mobile users in infrastructure-less zones. Table 2 presents a sample of the scenario dataset, listing the feature names, their data types, and representative entries used to initialize each node in the network simulation.

**Table 2** Sample of Scenario Dataset

| Feature Name | Data Type | Sample Entry |
|---|---|---|
| device_id | int64 | 1 |
| x_position | float64 | 271.513599 |
| y_position | float64 | 327.777799 |
| battery_level | float64 | 0.433033 |
| signal_quality | float64 | 0.784336 |
| success_rate | float64 | 0.483582 |
| device_type | object | phone |
| priority_tolerance | float64 | 0.673843 |
| buffer_capacity | int64 | 30 |
| uptime_ratio | float64 | 0.846589 |

During message delivery, detailed per-hop routing logs are recorded as csv files, capturing information such as TTL usage, delivery time, buffer state, success or failure, and full message paths. These logs enable fine-grained analysis of routing behavior and are used to extract training samples for machine learning models.

From the logs, four predictive datasets are derived: a success classifier dataset (Model A), TTL regressor dataset (Model B), delay regressor dataset (Model C), and best forwarder classifier dataset (Model D). All datasets include locally observable features such as TTL left, hop count, distance to target, buffer ratio, and device type. Labels are constructed to support binary classification, regression, and ranking, depending on the model's objective.

Synthetic data was chosen due to the logistical and privacy constraints of obtaining large-scale, labeled Bluetooth mesh logs in real deployments [19]. Our data generation process reflects realistic BLE-based behavior using AODV-inspired delivery logic and resource constraints under stationary node deployments. The resulting datasets enable scalable experimentation across a wide range of conditions while capturing both successful and failed routing attempts—critical for training reliable, context-aware models.

### 2.2. DEVICE MODEL AND NETWORK DYNAMICS

Each node in the mesh network is modeled as a Device object instantiated from scenario-specific CSV files. Devices possess attributes including position, battery level, signal quality, buffer capacity, uptime ratio, and device type. All nodes remain stationary during simulations, with no mobility modeled. Nodes maintain a buffer queue, simulate packet forwarding behavior, and dynamically discover neighbors within a fixed communication radius (50 meters). Public-key RSA encryption is optionally enabled per device for secure message exchange [20]. A capability score is computed as a weighted combination of battery level, signal quality, and historical success rate, contributing to delay estimation during transmission.

### 2.3. BASELINE AODV ROUTING WITH TTL CONSTRAINTS

The baseline routing mechanism follows a simplified AODV protocol constrained by a Time-To-Live (TTL) limit. A breadth-first search (BFS) is performed from the sender to discover a route to the receiver. If a path is found, each node updates its routing table accordingly. If no complete route is available within the TTL limit, the sender logs an early failed attempt using the top two nearest neighbors, simulating partial attempts. This mode serves as a fallback mechanism in the hybrid setup and provides a benchmark for comparison against ML-enhanced routing.

## 2.4. FEATURE ENGINEERING

The hybrid routing framework incorporates a context-aware and computationally efficient feature engineering process directly embedded within the simulation runtime. Eight features are dynamically extracted from the local state of the forwarding node and the message being routed. These features are selected to reflect real-time network conditions, node behavior, and delivery urgency, enabling intelligent routing decisions at each hop without requiring historical or global network knowledge.

The first feature, ttl_left, captures the number of hops remaining before a message expires. It is computed as the difference between the initial TTL assigned to the message and the current hop count, which is defined in Eq (1)

$$ttl_{left} = TTL_{initial} - hop\_count \quad (1)$$

Where $TTL_{initial}$ is the maximum number of hops permitted for the message, and $hop\_count$ is the number of hops the message has already taken. This feature helps the models identify time-sensitive messages and adapt routing behavior accordingly. The $distance\_to\_target$, calculates the Euclidean distance from the neighbor node to the intended recipient as in Eq (2)

$$distance\_to\_target = \sqrt{(x_2 - x_1)^2 + (y_2 - y_1)^2} \quad (2)$$

where $(x_1, y_1)$ and $(x_2, y_2)$ denote the positions of the neighbor node and receiver, respectively. This metric supports spatially informed forwarding. The $uccess\_rate\_origin$, is a measure of the forwarding node's past success in delivering messages, computed as Eq (3)

$$success\_rate\_origin = \frac{Successful\_deliveries}{Total\_Attempts} \quad (3)$$

This allows the model to favor historically reliable nodes. The Priority_tolerance, is a scalar value fixed per device, reflecting its designed capacity or willingness to process high-priority or urgent messages. This is used to differentiate forwarding roles in heterogeneous device environments. The *uptime_ratio*, tracks the node's operational availability and is calculated as Eq (4)

$$uptime\_ratio = \frac{Time\_Active}{Total\_Simulation\_Time} \quad (4)$$

This feature helps avoid unstable or intermittently offline nodes that could drop or delay packets. The *buffer_ratio*, measures the relative load on a node's message buffer, given by Eq (5)

$$buffer\_ratio = \frac{Buffer\_Used}{Buffer\_Capacity} \quad (5)$$

Nodes with high buffer usage are deprioritized to prevent overload and data loss. The *device_type_encoded*, represents the type of device (e.g., phone, sensor, relay) using ordinal encoding. This transformation converts categorical types into integer values compatible with tree-based models without introducing unintended bias.

All features are derived using current node and messaging metadata within the simulation environment. There is no historical aggregation, transforma tion, or derived feature interaction beyond normalization (applied only for Model D). In simulation, distance_to_target uses ground-truth coordinates; in practice it could be approximated from coarse localization or RSSI. Only lightweight local counters (uptime, past success) are used, without global or future information. This ensures the system remains lightweight, explainable, and suitable for deployment in resource-constrained mesh networks while still allowing robust real-time decision-making by the learning models [21].

## 2.5. MACHINE LEARNING MODELS

The hybrid routing framework employs four supervised machine learning models—Models A, B, C, and D—each serving a distinct predictive role in the decision-making process. These models are trained on contextual features extracted from simulation logs of AODV-based message deliveries and are integrated into the routing logic to improve adaptability, precision, and success rates in stationary Bluetooth mesh networks, with mobility left for future extension.

Model A is a binary classifier designed to estimate the probability of successful end-to-end message delivery when a message is forwarded through a specific neighbor. It was trained using a subset of context-aware features such as ttl_left, distance_to_target, and success_rate_origin. Several classification algorithms were tested, but XGBoost was selected as the final choice due to its proven ability to handle imbalanced data distributions, capture non-linear feature interactions, and provide high interpretability through feature importance scores. These properties are especially critical for mesh environments with heterogeneous conditions, where decision precision and transparency directly affect route reliability [22].

Model B predicts the expected TTL cost, which is the number of hops a message is likely to take to reach its destination via a specific neighbor. Features such as hop_count, ttl_left, and buffer_ratio were used. XGBoost was again chosen over linear regression and random forest models because of its robustness to multicollinearity and its superior ability to model complex, non-monotonic relationships between input features and hop count outcomes. The algorithm's regularization mechanisms also help reduce overfitting, which is beneficial in synthetic but variably structured datasets.

Model C estimates the delivery delay (in seconds) when forwarding through a specific neighbor. Although ensemble methods like XGBoost and SVR with RBF kernels demonstrated reasonable performance, Ridge Regression was ultimately selected for its simplicity, computational efficiency, and strong generalization in low-noise regression tasks. In delay prediction, where linear relationships often exist (e.g., between distance, priority, and time), Ridge Regression balances prediction accuracy and runtime feasibility which is key for real-time mesh simulation and deployment on constrained devices [23].

Model D acts as a top-k classifier that identifies the most suitable forwarder from a set of neighboring nodes. It uses a strict labeling scheme in which only the node that actually forwarded the message in each successful delivery is labeled as a positive instance. This approach requires the classifier to differentiate subtle contextual signals under significant class imbalance. Although XGBoost was initially tested, it exhibited low recall on the minority class. After hyperparameter tuning, a Random Forest classifier outperformed other candidates due to its resilience to overfitting, ease of interpretability, and strong class discrimination performance. Random Forest is particularly effective for datasets with categorical and continuous features, and its ensemble structure naturally balances bias and variance in decision boundaries [24]. Further Details of algorithm comparison, validation metrics, and final model selection are provided in Section 4.2.

## 2.6. MODEL FUSION AND DELIVERY LOGIC

Once individual predictions are generated by the four supervised models, a unified decision score is computed for each forwarding candidate in the Hybrid ABCD framework. This score integrates model outputs using a weighted linear combination that balances delivery reliability, route efficiency, and forwarder suitability. The formula is defined as Eq (6)

$$Score_{ABCD} = D + A - B - (\frac{C}{100})  \qquad (6)$$

Where, D is the forwarder probability from Model D (top-k classifier), A is the success probability from Model A (success classifier), B is the expected TTL usage from Model B (TTL regressor), C is the predicted delivery delay from Model C (delay regressor), normalized to a consistent scale. This formulation rewards candidates with high predicted reliability and suitability (via Models A and D), while penalizing those expected to introduce longer paths or delays (via Models B and C).

The specific fusion weights used in Equation (6) were empirically optimized through iterative testing across multiple simulation scenarios to achieve a practical trade-off between delivery success, latency, and efficiency. These values were selected based on observed system behavior rather than automated learning and are consistent with the scoring logic evaluated in Section 4.2.

To further enhance computational efficiency, Model D is optionally used first to shortlist the top-$k$ most promising neighbors based on forwarder suitability. For each candidate in this shortlist, the ABCD score is computed, and the neighbor with the highest score is selected as the next-hop forwarder.

At each hop, this hybrid decision logic is invoked in real time. If no suitable candidate surpasses a predefined threshold, or if Model D yields an empty shortlist, the system defaults to a classical AODV-based fallback mechanism to maintain delivery robustness.

All routing decisions including forwarder selection, model outputs, and message outcomes, are logged during simulation runtime. This enables detailed post-analysis and supports future refinement of the learning models. Algorithm 1 formalizes this decision-making process, suitable for real-time deployment in decentralized, stationary Bluetooth mesh networks.

---

**Algorithm 1:** Hybrid ML-AODV Routing Decision Using Model Fusion

**Input:**
  N: Set of current neighbor nodes

  M: Message metadata (TTL_initial, hop_count, sender, receiver, etc.)

  Features[n]: Extracted features for each neighbor n ∈ N

  models: {model_A, model_B, model_C, model_D} — Predictive models for success probability, TTL cost, delay, and forwarder ranking

  k: Top-k filtering threshold from Model D

  threshold: Minimum ABCD score for model-based forwarding

**Output:**
  Selected_forwarder ← neighbor node with highest ABCD fusion score

---

1. Predict Local Model Outputs:
   For each neighbor n ∈ N:
     score_A[n] ← model_A.predict_success_prob(Features[n])
     score_B[n] ← model_B.predict_ttl_cost(Features[n])
     score_C[n] ← model_C.predict_delay(Features[n])
     score_D[n] ← model_D.predict_proba(Features[n]) (class 1 = best forwarder) end for

2. Pre-select Candidates Using Model D:
     Candidates ← Top-k neighbors based on score_D (descending order)

3. Select Best Forwarder:
     Selected_forwarder ← Candidate with max score_ABCD[n]
     If no candidate exceeds threshold or selection fails:
       Selected_forwarder ← fallback_AODV_selection()

4. Return:
     Selected_forwarder

## 3. RESULTS AND ANALYSIS

The experiments were conducted on Jupyter Notebook, a web-based interactive computing platform, and the application was run in system with an 8 core CPU, 32.77GB RAM, and 8GB VRAM. Python is used as the programming language for implementing the model. The experimental details are outlined below:

### 3.1. EXPERIMENTAL SETUP

The proposed context-aware hybrid routing framework was evaluated through extensive simulations across ten synthetic yet realistic stationary scenarios. The analysis is divided into two major parts: (1) Evaluation of Local Machine Learning Models, and (2) Full Routing Strategy Comparison and Integration Analysis, covering classical AODV, ML-enhanced hybrid routing (ABC), and the full intelligent ABCD system. All evaluations are grounded in reproducible metrics including classification accuracy, regression error, and overall network-level performance indicators such as packet delivery ratio (PDR), average delay, TTL utilization, and hop count.

### 3.2. EVALUATION OF MACHINE LEARNING MODELS

To enable context-aware routing decisions at each hop, four supervised learning models were developed and trained using simulation logs generated under baseline AODV routing. These logs captured message trajectories, per-hop metrics, device conditions, and delivery outcomes, resulting in four distinct datasets with structured input features and ground-truth labels.

Model A, the Success Classifier, estimates whether a message forwarded through a given neighbor will reach its destination. It was trained using features such *as ttl_left, distance_to_target*, and *success_rate_origin*. Among all tested algorithms, XGBoost yielded perfect performance on validation data, achieving 100% accuracy and ROC AUC, outperforming Logistic Regression and KNN classifiers. This is summarized in Table 3.

**Table 3.** Performance Comparison of Model A (Success Classifier)

| Algorithm | Accuracy | F1 Score | ROC AUC |
|---|---|---|---|
| XGBoost | 1.0000 | 1.0000 | 1.0000 |
| Logistic Reg. | 0.9614 | 0.9758 | 0.9960 |
| KNN (k=5) | 0.9683 | 0.9803 | 0.9947 |

Model B, the TTL Regressor, predicts the number of hops required through a specific neighbor. This model helps the routing logic penalize longer or inefficient paths. XGBoost again achieved outstanding performance with an R² score of 1.000, as seen in Table 4. These perfect scores reflect deterministic patterns in the synthetic AODV-inspired logs; we verified no future information was leaked into training.

**Table 4.** Model B (TTL Regressor) Results

| Algorithm | RMSE | MAE | R² Score |
|---|---|---|---|
| XGBoost | 0.0124 | 0.0060 | 1.0000 |
| Random Forest | 0.0376 | 0.0103 | 0.9998 |
| Linear Regr. | 0.1551 | 0.1436 | 0.9960 |

Model C, the Delay Regressor, estimates the total expected delay in seconds if a neighbor is chosen for forwarding. The routing logic uses this model to favor faster routes. Although XGBoost and Random Forest showed good performance, Ridge Regression was selected for its simplicity and efficiency, achieving an R² of 0.9547, reported in Table 5.

**Table 5.** Model C (Delay Regressor) Results

| Algorithm | RMSE | MAE | R² Score |
|---|---|---|---|
| Ridge Regression | 10.46 | 8.05 | 0.9547 |
| XGBoost | 11.57 | 8.89 | 0.9445 |
| SVR (RBF Kernel) | 15.65 | 11.96 | 0.8985 |

Model D, the Best Forwarder Classifier, is trained using a strict labeling method where only the node that actually forwarded the message in a successful delivery is labeled positive. This enables learning of optimal next-hop decisions based on real-world network conditions. Although XGBoost initially underperformed with low recall and precision for positive class instances summarized in Table 6, a tuned Random Forest classifier dramatically improved results, reaching a recall of 99.68%, precision of 88.23%, and ROC AUC of 0.9965.

**Table 6.** Model D (Forwarder Ranking) Performance Comparison

| Model | Accuracy | F1 Score | ROC AUC | Precision (1) | Recall (1) |
|---|---|---|---|---|---|
| XGBoost (untuned) | 0.9217 | 0.2993 | 0.9170 | 0.8038 | 0.1839 |
| Random Forest (Tuned) | 0.9876 | 0.9361 | 0.9965 | 0.8823 | 0.9968 |

To ensure reproducibility, all final hyperparameters for each model were selected via grid search and cross-validation, as shown in Table 7.

**Table 7.** Final Tuned Hyperparameters of ML Models

| Model | Algorithm | Key Hyperparameters |
|---|---|---|
| Model A | XGBoost | subsample=1.0, n_estimators=200, min_child_weight=5, max_depth=3, gamma=0.2, colsample_bytree=1.0 |
| Model B | XGBoost | subsample=0.8, n_estimators=300, max_depth=8, learning_rate=0.1, colsample_bytree=1.0 |
| Model C | Ridge Regression | alpha=0.1 |
| Model D | Random Forest | max_depth=None, max_features='log2', min_samples_leaf=1, min_samples_split=2, n_estimators=200 |

### 3.3. ROUTING PERFORMANCE ANALYSIS AND MODEL INTEGRATION

To evaluate the real-world effectiveness of our context-aware hybrid routing system, we compared three routing strategies: (1) the classical AODV baseline, (2) the intermediate Hybrid ABC model using three predictive ML models, and (3) the full Hybrid ABCD model integrating four learning-based predictors into a unified decision-making score. All models were tested across ten distinct simulation scenarios, each representing realistic stationary mesh configurations with varied density and constraints.

### 3.3.1. BASELINE AODV CONFIGURATION

The AODV baseline implements a reactive route discovery protocol constrained by a Time-To-Live (TTL) limit. At each message initiation, a breadth-first search (BFS) is conducted from sender to receiver, terminating once the destination is reached or TTL is exhausted. Routing tables are updated upon successful delivery. In failure cases, up to two partial paths are simulated using the nearest neighbors to mirror real-world failed attempts. This model provides a consistent, low-overhead benchmark, but exhibits limitations under link variability and buffer congestion, often selecting poor intermediate nodes or exceeding TTL budgets.

### 3.3.2. HYBRID ABC ROUTING

In the Hybrid ABC model, machine learning enhances the traditional AODV protocol by evaluating each potential forwarding neighbor using a composite scoring mechanism that incorporates three predictive insights: delivery success probability, TTL cost, and estimated delay. These correspond to the outputs of Models A, B, and C as defined in Section 3.6. To ensure output consistency, the delay prediction is normalized. While a general scoring formula was introduced earlier as Eq (6), the version below defined as Eq (7) reflects empirically tuned weights derived from simulation experiments:

$$Score_{ABCD} = 0.5 \cdot A - 0.25 \cdot B - 0.25 \cdot (\frac{C}{100}) \tag{7}$$

All neighbors are scored using this formula, and the one with the highest value is selected as the next forwarder. While this improves routing decisions by factoring in multiple quality metrics, it may still admit suboptimal candidates due to prediction variance.

### 3.3.3. HYBRID ABCD ROUTING

The Hybrid ABCD model extends this approach by directly incorporating Model D's prediction. With the suitability of each neighbor as a forwarder into the scoring function. Rather than using D solely for top-k pre-filtering, it is weighted alongside A, B, and C in the final decision. The resulting tuned scoring formula is presented in Equation (8)

$$Score_{ABCD} = 0.4 \cdot D + 0.4 \cdot A - 0.1 \cdot B - 0.1 \cdot (\frac{C}{100}) \tag{8}$$

This integration improves the routing engine's ability to select high-quality paths under varied network conditions. In practice, the system may first shortlist top-k candidates using Model D and then apply Equation (8) to select the best forwarder. This two-stage selection enhances both computational efficiency and decision robustness, particularly in congested networks with constrained resources.

The effectiveness of this integration is clearly demonstrated in the results summarized in Table 6. The Hybrid ABCD model achieves a delivery success rate of 99.97%, significantly outperforming both the Hybrid ABC (88.87%) and the baseline AODV (76.65%) across all ten scenarios. Moreover, ABCD consumes more TTL (less TTL remaining), with an average TTL left of 5.51 per message, compared to 7.09 in Hybrid ABC and 6.97 in AODV. This reflects deeper routes that improve reliability at the cost of higher delay. While the average delivery delay in ABCD (140.46 seconds) is higher than in Hybrid ABC (103.61 seconds) and AODV (58.53 seconds), this trade-off is acceptable in emergency scenarios where message delivery reliability outweighs strict latency requirements [25]. Many real-world disaster communication use cases can tolerate delays in the order of minutes, provided messages are eventually delivered with high confidence [26][27]. This delay increases results from deeper routing paths (average of 4.49 hops in ABCD vs. 2.91 in Hybrid ABC and 2.53 in AODV), which are necessary to ensure successful delivery in stationary, decentralized environments [28][29].

Table 8. Network-Level Routing Metrics Across Models

| Mode | PDR (%) | Avg TTL Left | Avg Delay (s) | Avg Hops |
| --- | --- | --- | --- | --- |
| Baseline | 76.65 | 6.97 | 58.53 | 2.53 |
| Hybrid ABC | 88.87 | 7.09 | 103.61 | 2.91 |
| Hybrid ABCD | 99.97 | 5.51 | 140.46 | 4.49 |

Figure 2 further visualizes the per-scenario delivery performance, confirming that Hybrid ABCD consistently delivers nearly all messages in every test case, while both ABC and AODV show significant variability depending on scenario complexity. This reinforces the ABCD model's robustness and generalization capability across diverse network conditions. Across 10 scenarios, Hybrid ABCD improved mean PDR by +23.3 percentage points over AODV (95% CI ±2.1pp; Wilcoxon signed-rank test $p<0.01$).

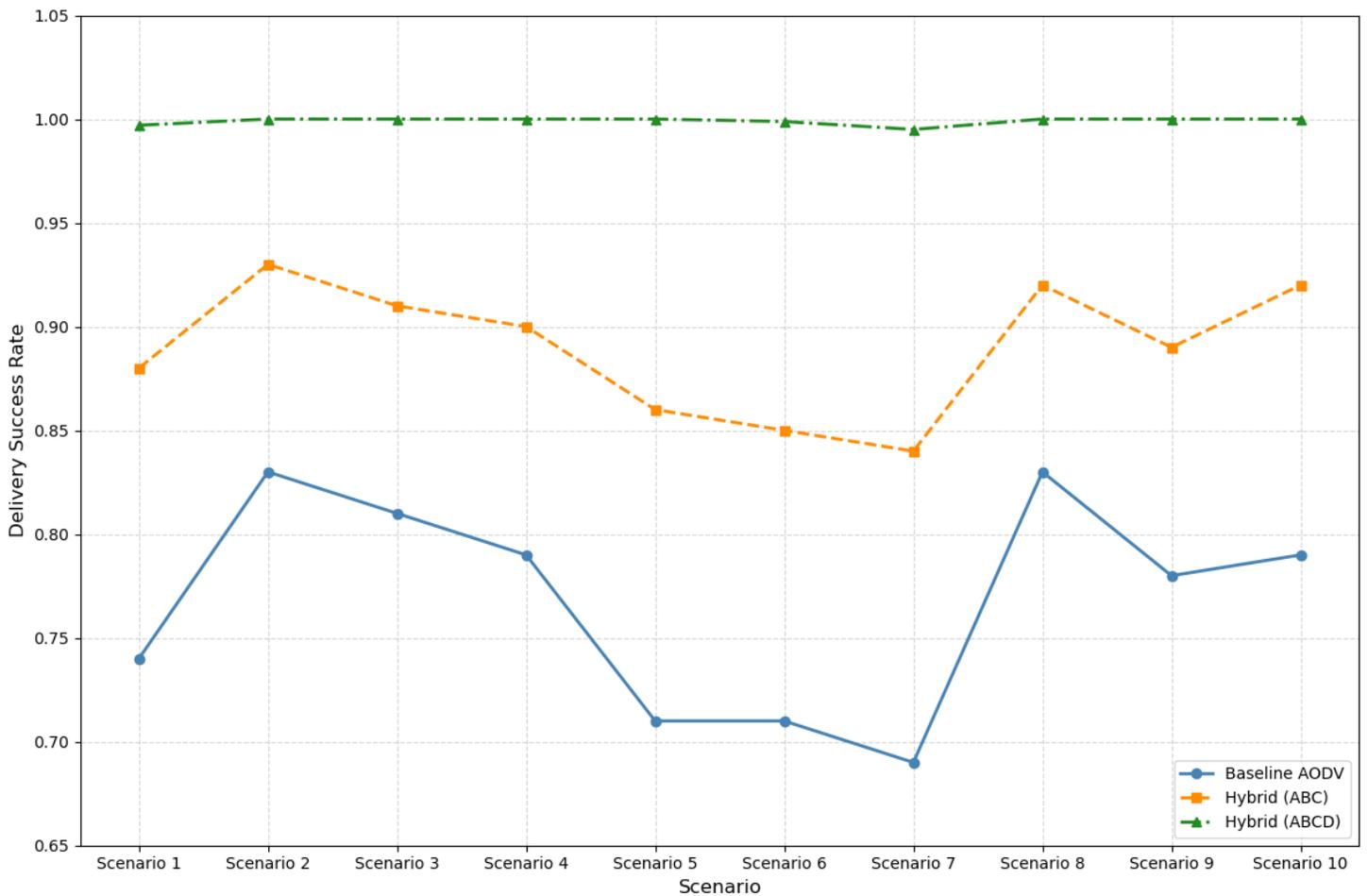

Figure 2. Per-Scenario Delivery Success Rate by Routing Mode

## 3.4. DISCUSSION AND FUTURE WORK

The proposed Hybrid ABCD routing framework demonstrates strong improvements in delivery reliability, adaptability, and routing intelligence within stationary Bluetooth mesh networks. By integrating four lightweight supervised models into a unified decision system, the framework achieves a 99.97% packet delivery ratio, significantly outperforming classical AODV (76.65%) and the intermediate Hybrid ABC model (88.87%) across ten varied stationary scenarios. Model D, which ranks neighbors based on contextual suitability, notably enhances decision quality under constrained network conditions.

The multi-objective scoring function balances delivery success, TTL efficiency, and delay estimation, producing more purposeful routing paths. Although ABCD incurs a higher average delivery delay (140.46 seconds), this trade-off is acceptable in delay-tolerant scenarios where reliability is the priority [25][26][27]. Its reliance on local features and real-time inference ensures scalability without centralized control or historical data.

Nonetheless, several limitations guide future work. First, all evaluations used synthetically generated data based on AODV-like logic. While structured and controlled, these simulations may not capture environmental factors such as signal interference, hardware variability, or stochastic channel effects. Moreover, this study was restricted to stationary deployments; extending the framework to mobile BLE mesh networks is an important direction for future validation [1]. Future validation on BLE-compatible hardware (e.g., nRF52840 dongles) will also be essential to assess generalizability.

Additionally, while the current scoring function uses fixed weights, future studies may explore adaptive weighting through reinforcement learning or meta-fusion strategies, enabling real-time adjustment under evolving conditions. Two planned but unimplemented features warrant clarification. A fifth model (Model E), designed for energy-aware routing, was conceptualized but not included in this version. Likewise, RSA encryption was present in the simulation layer as a placeholder, but not functionally active in routing or security evaluation. Both will be addressed in future iterations to enhance energy optimization and privacy support.

Finally, broader extensions could include support for intermittently connected friend-nodes and federated learning techniques to refine models without raw data exchange, enhancing privacy and robustness in decentralized networks..

## 4. CONCLUSION

This study introduced a hybrid intelligent routing framework for Bluetooth mesh networks, combining classical AODV with four supervised machine learning models to optimize next-hop forwarding under resource-constrained conditions. By integrating predictions for delivery success, TTL cost, delay estimation, and forwarder suitability into a unified scoring mechanism, the proposed Hybrid ABCD model achieved near-perfect delivery reliability (99.97%) across diverse stationary simulation scenarios. Compared to both baseline AODV and the intermediate Hybrid ABC variant, the ABCD model demonstrated superior adaptability and decision quality, particularly under constrained and decentralized network environments.

The framework's reliance on locally available features and interpretable models ensures real-time operation, scalability, and deployment feasibility. While the current evaluation is simulation-based, future work will focus on real-world testing, adaptive fusion strategies, and energy-aware extensions to support broader use cases. Moreover, extending the framework from stationary to mobile BLE mesh networks remains a key direction to validate robustness under dynamic conditions.

Overall, this study demonstrates that intelligent routing in decentralized Bluetooth mesh networks can be achieved through a multi-model fusion strategy, offering a novel, scalable, and interpretable alternative to deep learning or heuristic-based approaches.